# Self-Organized Networks with Long-Range Interactions:

# Tandem Darwinian Evolution of α and β Tubulin

J. C. Phillips

Physics Dept., Rutgers Univ., Piscataway, N. J., 2019-19414P

## Abstract

Cytoskeletons are self-organized networks based on polymerized proteins: actin, tubulin, and driven by motor proteins, such as myosin, kinesin and dynein. Their positive Darwinian evolution enables them to approach optimized functionality (self-organized criticality). Our theoretical analysis uses hydropathic waves to identify and contrast the functional differences between the polymerizing α and β tubulin monomers, which are similar in length and secondary structures, as well as having indistinguishable phylogenetic trees. We show how evolution has improved water-driven flexibility especially for α tubulin, and thus facilitated heterodimer microtubule assembly, in agreement with recent atomistic simulations and topological models. We conclude that the failure of phylogenetic analysis to identify functionally specific positive Darwinian evolution has been caused by 20th century technical limitations. These are overcome using 21st century quantitative mathematical methods based on thermodynamic scaling and hydropathic modular averaging. Our most surprising result is the identification of large level sets, especially in hydrophobic extrema, with both thermodynamically first- and second-order scaled water waves. Our calculations include explicitly long-range water-protein interactions described by fractals. We also suggest a much-needed corrective on drug development costs.

## Introduction

Self-organized networks are off-lattice structures described in many ways [1]. Although the structure of the Internet affects us daily [2], even more important are protein networks of living matter. The self-assembling proteins of cell-shaping cytoskeletons have been studied extensively [3,4]. Broadly speaking, the cytoskeletons of eukaryotes are composed of polymerized filaments and motors, while the motors are absent from prokaryotes [5]. The filaments exhibit striking elastic and electrical properties [6,7]. Self-organization of cytoskeleton proteins.has been parameterized [8], and self-organization exhibits many scale-free properties [9].

The alpha-beta tubulin heterodimer is one of the two major structural subunits of cylindrical microtubules, which are cytoskeletal filaments that are essential for intracellular transport and cell division in all eukaryotes. Their dual polymeric nature is an ideal subject for Darwinian evolutionary analysis; it involves several levels of thermodynamic scaling theory, previously applied to actin, a simpler single polymeric component of the cytoskeleton [10]. A similar discussion has been given for tubulin [11]; this earlier discussion led to an anonymous suggestion that more could be done in a lengthier analysis, and that is given here. The central new result, which deserves separate publication, is a clear-cut separation of thermodynamically distinct first-and second order effects that occur at different water wave lengths ($W* = 25$ in [11], and $W* = 33$ here). Although this distinction is fundamental to thermodynamics, it is not readily accessed in molecular dynamics simulations. Here it is easily obtained by comparing results obtained with the firsty-order KD scale and the second-order MZ scale [11,12].

The general method used here is only ten years old, and is little known. It is based on a general thermodynamic concept, self-organized criticality (SOC); see [13] for a popular recent review, which unfortunately omitted fractals. SOC has proved to be a powerful concept for analyzing the evolution of protein function from sequences alone, abundantly available in the genomic era. SOC originated in 1987 as an extension of fractal geometries to thermodynamic systems in the vicinities of instabilities [11]. Many physical systems exhibit power-law distributions over limited ranges (hence the enduring popularity of log-log plots), and power-law distributions are the characteristic feature of the modern theory of phase transitions near a critical point. SOC is a methodology that attempts to explain why so many complex systems exhibit scale-free power-law distributions [9] and appear to be "accidentally" located near critical points. It is argued that the critical points are dynamical fixed points ("tipping points") towards which the system evolves without tuning external parameters. The critical points are extrema in some property (or properties) with respect to which the system has been nearly optimized, especially with respect to long-range, highly cooperative interactions. SOC explains simply and quite generally the power-law distributions that are observed in many complex self-organized networks over 6 decades.  It formalizes common experience as described by the law of diminishing returns.  It is reviewed further in our earlier companion article on actin [1].  Actin polymers are also part of the cytoskeleton and are only 16% smaller.

The ultimate phylogenetic goal, of establishing the power of Darwinian selection for improving protein function, has not been attained [12]. However, efforts to quantify molecular clocks, which were begun already in the 1960's by Pauling and others, have yielded positive results. More generally, there are many difficulties in phylogenomics, and "more sequences are not enough" [13].

Phylogenetics counts numbers of identical or similar amino acids at specific sites using BLAST, and it is severely limited by the restriction to single sites.  There is a thermodynamic alternative to the single site methods, which has Darwinian selectivity as an implicit feature, as corroborated by the identification of universal self-organized criticality in the solvent-accessible surface areas (SASA) of > 5000 protein amino acid segments from the modern Protein Data Base  [14].  The lengths of the small segments L = 2N + 1 varied from 3 to 45, but the interesting range turned out to be M< = $9 \leq L \leq 35$ = M>.  Across this range Moret and Zebende [14] found linear behavior on a log-log plot (a power law, hence self-similar or scale-free) for each of the 20 amino acids centered on a given segment

$$\log SASA(L) \sim const \ - \ \Psi(aa) \ \log L \quad \quad ( \ 9 \leq L \leq 35)$$

Here $\Psi(aa)$ is a hydropathicity parameter (technically a fractal).  It arises because the longer segments fold back on themselves, occluding the SASA of the central aa.  The most surprising aspect of this self-similar folded occlusion is that it is nearly universal on average across the proteome, and almost independent of the individual protein fold.   This is a dramatic demonstration of the power of Darwinian selectivity involved in aqueous shaping of globular proteins.

The simplest example of the power of water wave scaling is the monotonic evolution of the centrosymmetric secondary structure of  Hen Egg White from chickens to humans [15]; many other examples are discussed in detail elsewhere [16].  Moreover, the segmental character of the new scale [14,15] has a Darwinian echo: for each protein family one can identify a nearly optimized sliding window width W*, over which Ψ(aa) is well averaged to maximize moderate evolutionary improvements in standing water waves; this averaged profile is denoted by Ψ(aa,W*). Profiles of Ψ(aa,W*) display the functional features that are being optimized by evolution, often involving modular (segmental) exchange [16].

Like actin, tubulin has evolved very little, so we look for a reference species which is the best starting point for measuring eukaryotic evolution. A comprehensive search for conserved elements in vertebrate genomes found roughly 3%-8% conserved elements of the human genome and substantially higher fractions of the more compact fruit fly (37%-53%), round worm (18%-37%), and single-cell fission yeast (47%-68%), so we use the latter as our reference species [17]. Yeast studies have led to the discovery of genes involved in fundamental mechanisms of transcription, translation, DNA replication, cell cycle control, and signal transduction. However, since the divergence of the two species approximately 350 million years ago, fission yeast appears to have evolved less rapidly than budding yeast, so that it retains more characteristics of the common ancient yeast ancestor, causing it to share more features with metazoan cells [18].

Hydropathicity scales measure the roughness of a globular protein in terms of the shape and variable density of its covering water film. The most important feature of a globular shape is its hydropathic compactness or roughness, which is determined by its curvatures near its core or surface extremities. The average curvatures are related to the variances (mean of the square minus square of the mean) of Ψ(aa,W). It is easily seen from examples that $\text{Var}(\alpha + \beta) \geq (\text{Var}\alpha + \text{Var}\beta)/2$. Since tubulin α and β have co-evolved as heterodimers, in thermodynamic scaling one can avoid ambiguities associated with trying to link their tandem evolution [19,20] by studying the evolution of $\text{Var}(\alpha + \beta)$.

Another tool useful in quantifying protein evolutionary differences is level sets of Ψ(aa,W) profile extrema. Level sets have been developed by mathematicians (see Wikipedia article on Level-set method) primarily for image analysis , but they have turned out to be a striking feature of tubulin profiles. Generally they are a useful tool for studying hydrodynamically the kinetics of phase transitions, including protein functions [16, 21-25]. One can divide the protein into dynamically functioning domains centered on hydrophobic extremal pivots, with edges at hydrophilic extrema hinges. Such level sets can facilitate synchronized domain motion. The effectiveness of one-dimensional level set concepts for tubulin may be related to microtubules' one-dimensional axes and its two-dimensional cylindrical surfaces. Mathematically oriented readers will find "simple" explanations of level sets and their comparative computer science advantages online, for instance, under "Level Set Methods: An initial value formulation".

**Results**

To quantify thermodynamically the significance of protein region curvatures with water, one should compare results obtained from the 2007 MZ fractal scale [14] (which implicitly describes thermodynamically second-order phase transitions), with those obtained from first-order protein unfolding measured by enthalpy changes from water to air (1982 KD $\Psi$ scale [15]; this is also the most popular $\Psi$ scale, which we call the standard scale). According to BLAST, the positive similarity sites for human and yeast tubulin are 88% ($\alpha$) and 89% ($\beta$), so we expect similar evolutionary patterns. However, large differences in domain separations were found in [11] using W* = 25, and here we explore further large difference in level sets with W8 = 33.

We begin with the $\alpha$ $\Psi$(aa, W*) profiles for the KD scale (Fig. 1) The monomer tubulin structures were divided into three d = 3 domains, N terminal, Intermediate, and acidic C terminal, 382-440 [27-29]. The KD profile in Fig. 5 matches these three domains very well, which is a simple test for the choice of W*. The human profile is more hydrophobic than the yeast profile, especially near site 300 (center of the intermediate domain), where a hydrophilic yeast hinge has been made nearly hydroneutral and more compact in the human sequence. The reduction in softening water film density near the central site 300 provides additional stability and longer life for mechanically more active human cells compared to passive yeast cells.

The level sets of both hydrophobic and hydrophilic extrema in Fig. 1 are striking features of this profile (KD scale with W* = 25), which are absent with the MZ scale with W* = 25 [11]. Is this a failure of the MZ scale, or merely a poor choice for W*? We searched our $\Psi$(aa,W) matrix (which has only a few thousand elements) and found that W* = 33 with the MZ scale gives excellent results for level sets (see Fig. 2). These level sets are discussed in more detail in the Figure captions; note that they are present in human profiles and absent from yeast profiles.

While there appeared to be little difference between α and β monomers in early static structure studies [16-18], more recent atomistic simulations of rearrangements upon hydrolysis of template pig (99% human sequence) structures extracted from ~ 100 mostly human structures have shown large differences in elasticity [19]. Given that hydrolysis is thermodynamically first-order, we expect to see these differences especially clearly by comparing the KD profiles of α (Fig. 1) and β (Fig. 3). Fig. 4 of [19] exhibits large hydrolysis displacements for α tubulin, and small displacements for β tubulin. Moreover "Remodeling of longitudinal dimer contacts is coupled to {long range} conformational changes in α-tubulin … The observed changes at the interdimer contact, around the E-site, are accompanied by internal rearrangements of the tubulin dimer involving the intermediate and C-terminal domains of α-tubulin …upon hydrolysis the α-tubulin intermediate domain within the microtubule undergoes a shift similar to that reported for the straight to bent transition".

Comparing Figs. 1 and 3, we see that α-tubulin's much wider range of hydropathicity with the KD scale is consistent with its observed much larger hydrolysis displacements [19]. Stacking of arenes (aromatic rings like benzene) is commonplace in biomolecules [20]. As noted in [21], in the N-terminal domain of α-tubulin. Phe 87, Phe 138, Phe 169, Phe 202 are stacked. Three of these Phe sites lie near $W^* = 25$ profile extrema, and the extrema near Phe 87, 169 and 207 are all more extreme in the KD profile (Fig. 5) than in the MZ profile (Figs. 6 and 7). These extrema mark turning points (hydrophobic pivots or hydrophilic hinges) for conformational motion, and are consistent with curved tubular cylinders shaped by long-range (allosteric) interactions. Figs. 5 and 8 show Darwinian evolution from mechanically passive yeast cells to human mechanically active tubulin. Compared to β tubulin, evolution has improved water-driven flexibility especially for α tubulin, and thus facilitated microtubule assembly [21]. The close agreement here between two quite different theoretical methods - thermodynamic scaling and atomistic molecular dynamics – is novel, and enhances their mutual significance.

The choice of $W^* = 33$ for the MZ scale yields better results not only for level sets of tubulin α, but also for tubulin β evolution from yeast to human, as shown in Fig. 4. The differences between Figs. 3 and 4 reflect the differences between short- and long-range interactions. Similar but smaller differences are present for other species, such as trout (not shown).

The sliding window W reveals a difference from the separated α and β profiles in the heterodimer profile near their boundary, as shown in Fig. 5. Both the KD and MZ polymer profiles with W* = 21 of yeast and human are correlated at 83%, roughly the average of the BLAST identities (77%) and positives (88%). Thus the overall W α,β correlations are no more informative than BLAST with regard to yeast-human evolution. The differences between the KD and MZ (W* = 25) profiles in terms of hydropathic stability, identification of domain edges, and even arene stacking to promote tubular stabilization, all favor the KD scale.

**Discussion**

The results obtained here concerning the hydropathic differences between α and β tubulins not only confirm the recent simulated results based on pig structures [19], but also show how Darwinian evolution from yeast to humans has refined and improved these differences. The most recent general review emphasizes the importance of heterodimers [21]. It resolves the paradox that while cylindrical filaments "appear to have lower degrees of freedom than individual dimers, the entire system has increased entropy due to the increased accessible states of the water molecules. When tubulin dimers are in solution, the hydrophobic patches that mediate dimer-dimer interactions are exposed. Water molecules have reduced degrees of freedom around hydrophobic patches, but are released when dimers bind to each other". (Of course, water is softer than the protein peptide-amino acid backbone; its boiling point is lower than the backbone melting point.) This process would be furthered by staggering α and β dimers on adjacent chains (rather than aligning them, as they suggested in their Fig. 1A [21]), yielding fully cylindrical strain-free matching [7].

Alternating adjacent stiff (hydrophobic) and flexible (hydrophilic) elements provides stability and adaptability, as in macroscopic bone and cartilage structures. They also create topological edge phonons which are stable against thermal fluctuations ("topological protection") [22,23]. Here again independent and parameter-free theories are in excellent agreement on an important feature of tubulin heterodimers that has so far eluded experiment. It is striking that [19]'s simulated results are rich in detail, obtained by combining 3-d structures with the enormous power of modern computers, while our thermodynamic results here take full advantage of the enormous 21$^{st}$ century genomic 1-d sequential data base [35]. Our calculations are much more economical and can be done with EXCEL on a laptop. The underlying cause of the dimeric stability of many proteins is compactly described by topological models of edge states in crystals [22,23] apply equally well to protein water films.

The $\alpha - \beta$ binding occurs through the amphiphilic (linear cascade) plus ends, as shown in Fig. 5. The length of the longer $\alpha$ amphiphilic fragment is about 40 aa. The amyloid $\beta$ fragment is also amphiphilic, with a length of about 30 aa (see Fig. 4 of [21]). Both these lengths are in the range $W^* \sim 25 - 33$, and are much larger than the standard lengths $W \sim 1$ used in phylogenetics [24]. In a different direction, it is quite possible that the ubiquitous power law tails recently discovered in coevolving pairs ($W = 2$ here) of sequence positions [25] are vestiges of the power laws discovered by Moret and Zebende [14]. These become functionally most important for $W \geq 7$, for example, in actin, where functional singularities are found at $W = 9, 21,$ and $35$ [10].

While monomer actin filaments have evolved little from prokaryotic algae to eukaryotes, prokaryote algae tubulin profiles are qualitatively different from eukaryote tubulin profiles. Presumably the difference arises from eukaryotic tubulin transportation involving nuclei and organelles [26]. The transport presumably occurs by ratcheting, which is thermodynamically first order as it involves unfolding the transported proteins [27].

At the cellular level microtubules organize the cell interior through pushing and pulling on its membranes in different ways for different cells [28]. These differences reflect thermodynamic differences involving mixed first-and second-order interactions, as reflected in the differences between the KD and MZ scales for $\alpha$ and $\beta$ proteins separately (Figs. 1/2). However, these differences largely disappear for connected $\alpha + \beta$ proteins (Fig. 5), which share a common extremum near $W = 25$. This value of $W$ is typical for membrane proteins [29]; it reflects long-range interactions.

There is growing evidence that the universality of the power law fit to SASA reported in [14] is also present in other correlations in the sequence data of known interaction partners [30]. These interactions often involve hydrophobic "hot spots", so further quantification could result from combining these methods with thermodynamic scaling. One would expect that globular proteins, shaped by hydropathic interactions, would exhibit analytic features reflecting level set dynamics, as discussed in mathematical models of phase transition kinetics [11]. There is an excellent Wiki on the history and applications of self-organized criticality, which shows many simple applications of the concept popularized by Mandelbrot and Bak [13]. Synchronizing network domain interactions (Fig. 2) is also an attractive way of enhancing neural network activity [31-35].

The costs of developing new drugs are increasing steeply and currently exceed $2 billion/new drug [36]; ~ 80% of these new drugs involve proteins. During development, many candidate proteins with known amino acid sequences (but not structures) are screened without benefit of the high-throughput methods that worked so well for easily prepared small-molecule drugs. Screening would be greatly simplified if preliminary mutated protein property measurements could be correlated with available sequences. There is a close parallel between natural Darwinian protein selection and screening candidate proteins for desired properties, but the inability of conventional methods to quantify Darwinian evolution [37] means they cannot be used in drug research. Such correlations are achievable swiftly and accurately with the thermodynamic water wave methods described here (for example, amyloids, HPV vaccine, aspirin, etc. [29]). These quantitative methods probably require much less effort and expense than is currently being spent on "quant" screening of financial markets [38].

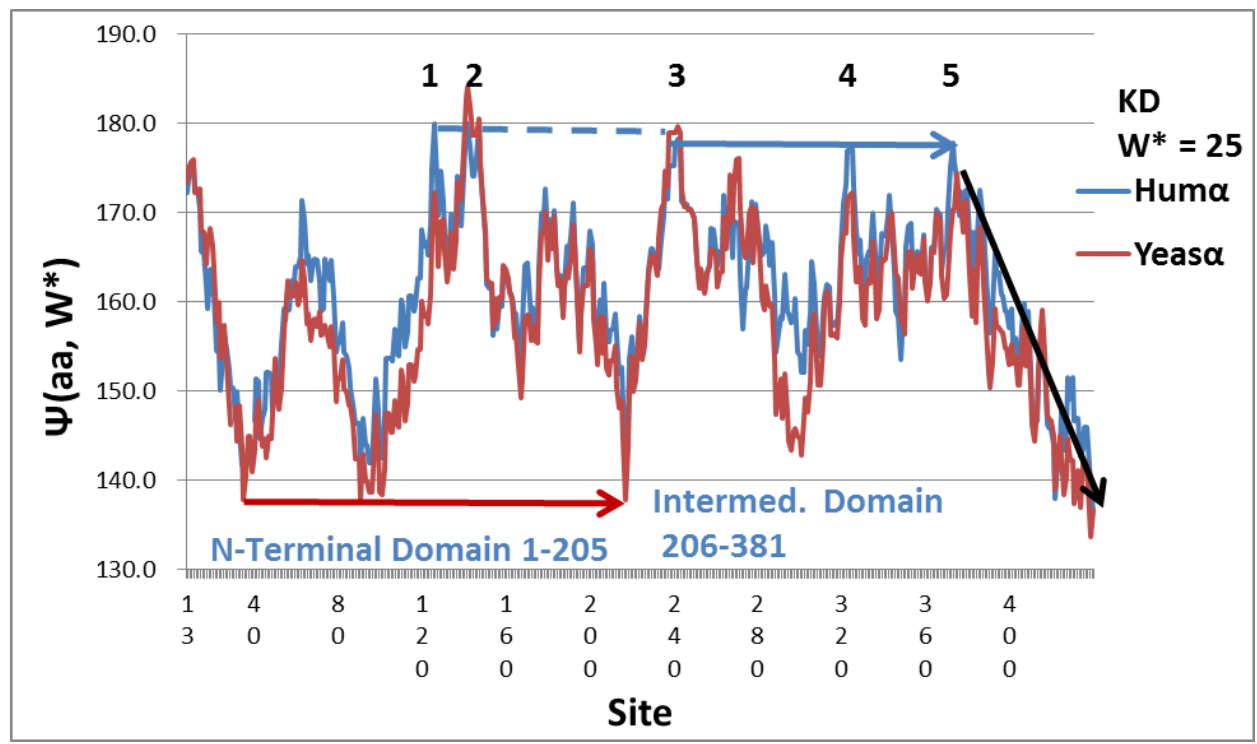

Fig. 1. The α tubulin W* = 25 profiles for human and yeast are quite similar (KD scale). Note the match between profile extrema and the tripartite domain edges identified structurally (see text). Also impressive are the three extremely level (Ψ ave. deviation. 0.1) human hydrophobic extrema (3-5) in the intermediate domain. The dashed line shows that two human peaks (1,2) in the N-terminal domain are also nearly level with these three (Ψ higher by only 1). The amphiphilic linearity of the 382-440 C-terminal domain is emphasized by the black arrow. Note also the water-driven flexibility of the N-terminal domain associated with its three deep and nearly level [36-38] hydrophilic hinges. Note the upward profile shifts of human above yeast around site 210 by about 10% of the profile range, and around site 300 by about 35% of the profile range. These shifts stabilize mechanically active human cells, which are subjected to greater stresses than passive yeast cells.

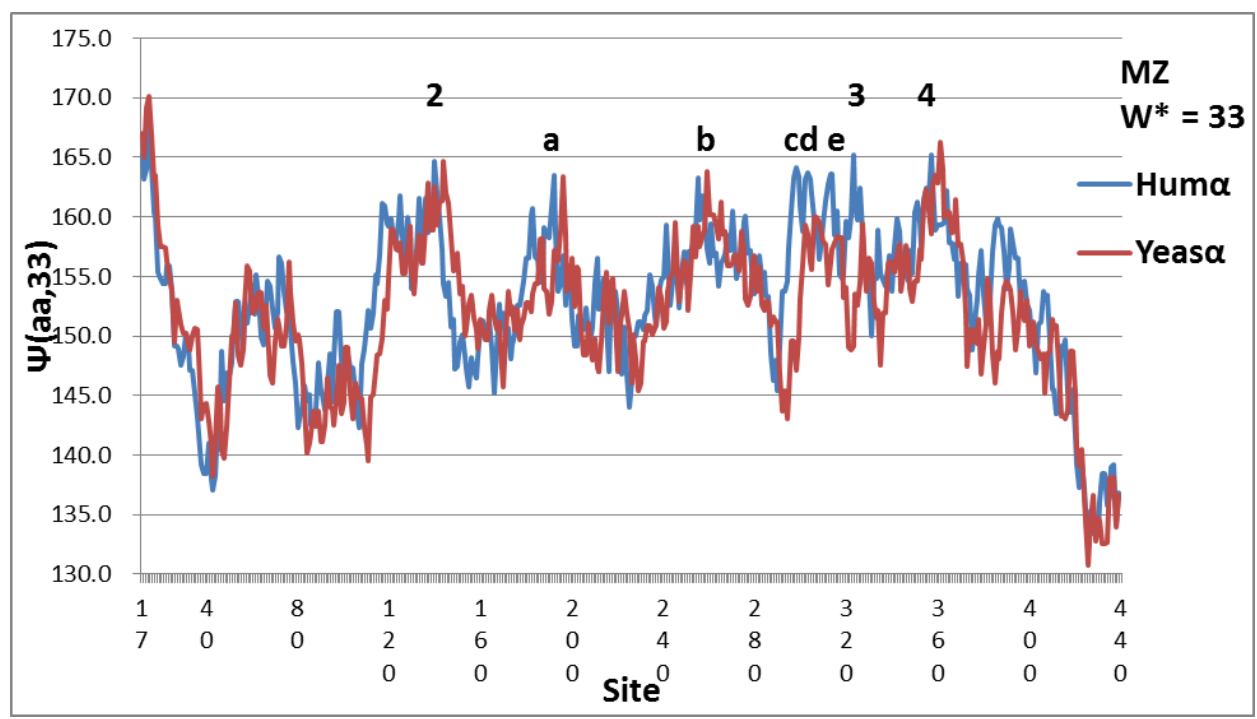

Fig. 2. The MZ scale with W* = 33 shows eight level peaks in sets of 3 and 5 for Human α. The sets are well separated at 165.0 (2) and 163.4 (1) respectively. Therefore the choice W* = 33 for the MZ scale is at least as effective in defining level sets as W* = 25 was for the KD scale in Fig. 1. The BLAST alignment of the Human and Yeast α sequences has a gap of 5 amino acids at human 40, which produces the profile offsets. The offsets facilitate comparison of Human and Yeast extrema. The differences are usually small, but between 300 and 330 they are large, with Humα much more hydrophobic. This is similar to the KD W* = 25 Human and Yeast differences, while here the Human peofile is refined and includes extra peaks c, d,e and 4. These extra hydrophobic peaks both stabilize (because more hydrophobic) and increase flexiblity (because the additional level peaks facilitate synchronized motion) of the Intermediate domain, and couple it (through 2 and a) to the N-terminal domain. At the same time, the larger value of W* = 33 and use of the MZ scale does not give the successful domain separation shown in Fig. 1 for W* = 25 with the KD scale.

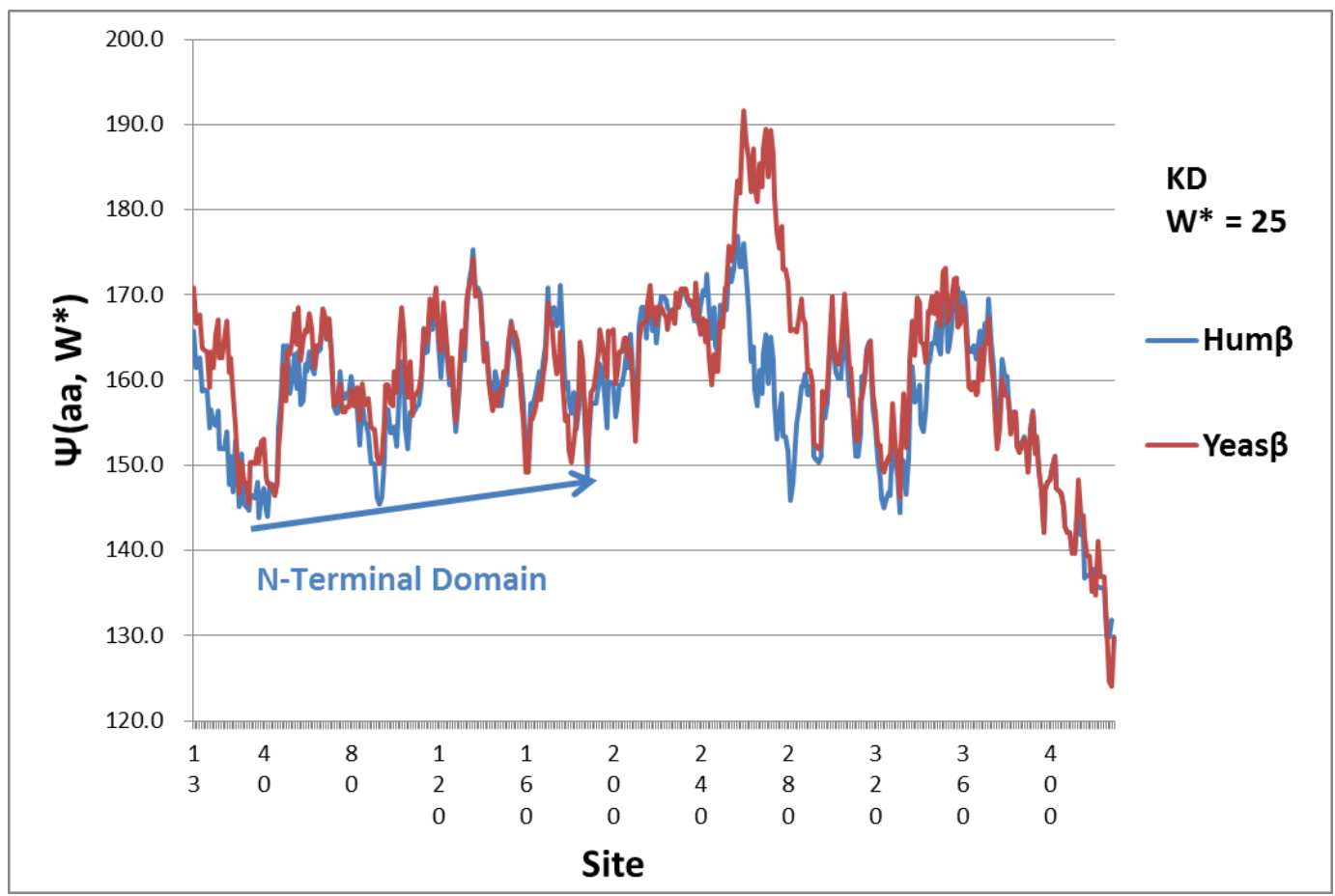

Fig. 3. Comparison of the KD β profile here with the KD α profile (Fig. 1) shows important evolutionary differences. Most striking is the disappearance of the very large hydrophobic β peak 260-280 from yeast to human, making human more flexible. In human β the ranges of both the N terminal and intermediate domains has narowed by about 40% compared to the α range. For example, the hydrophilic hinges of the N-terminal domain here average Ψ(aa,25) are around 150, whereas in Fig. 5 for α tubulin, they were all below 140.

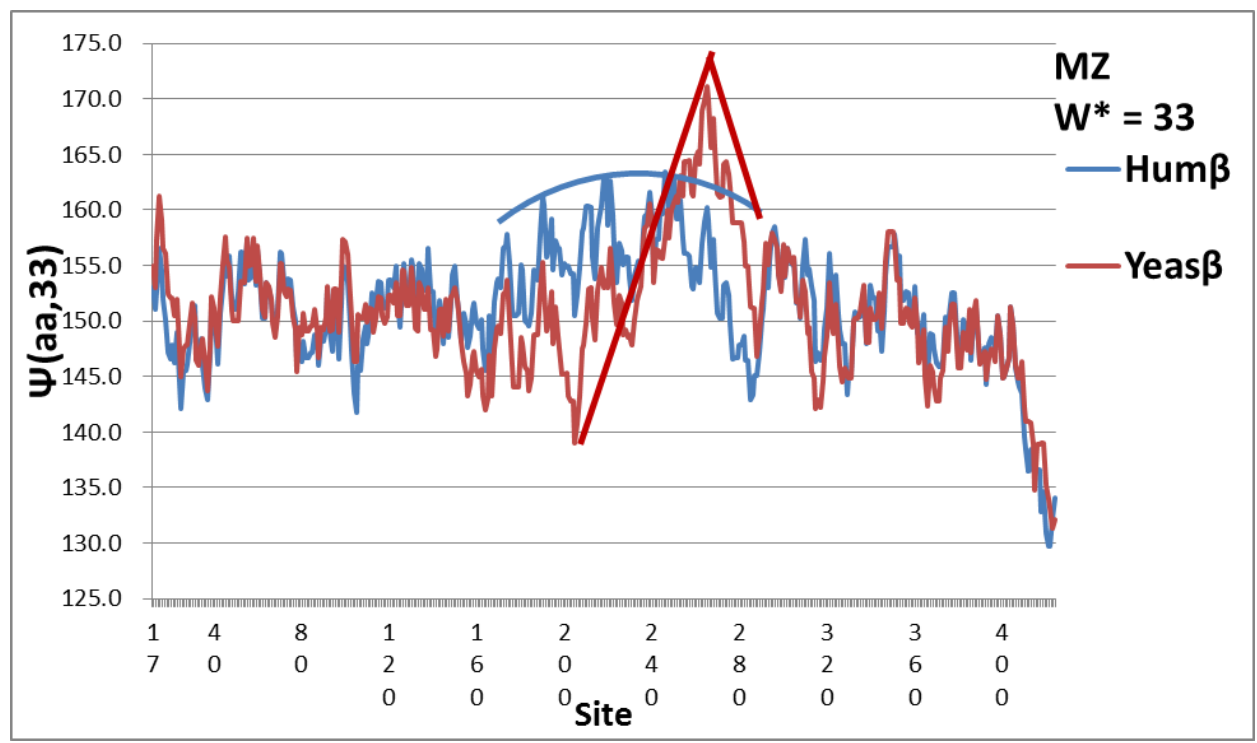

Fig. 4. Comparison with Fig. 3 shows that human β MZ profile appears to have nearly erased the site 206 N-terminal/Intermediate domain boundary, compared to yeast. Whereas yeast had a centered principal hydrophobic peak near site 275 (resembling an inverted "V"), human β has double hydrophobic peaks near sites 200 and 250, with stronger stabilzation with and improved allosteric contact between the Intermediate Domain and the N-Terminal domain.

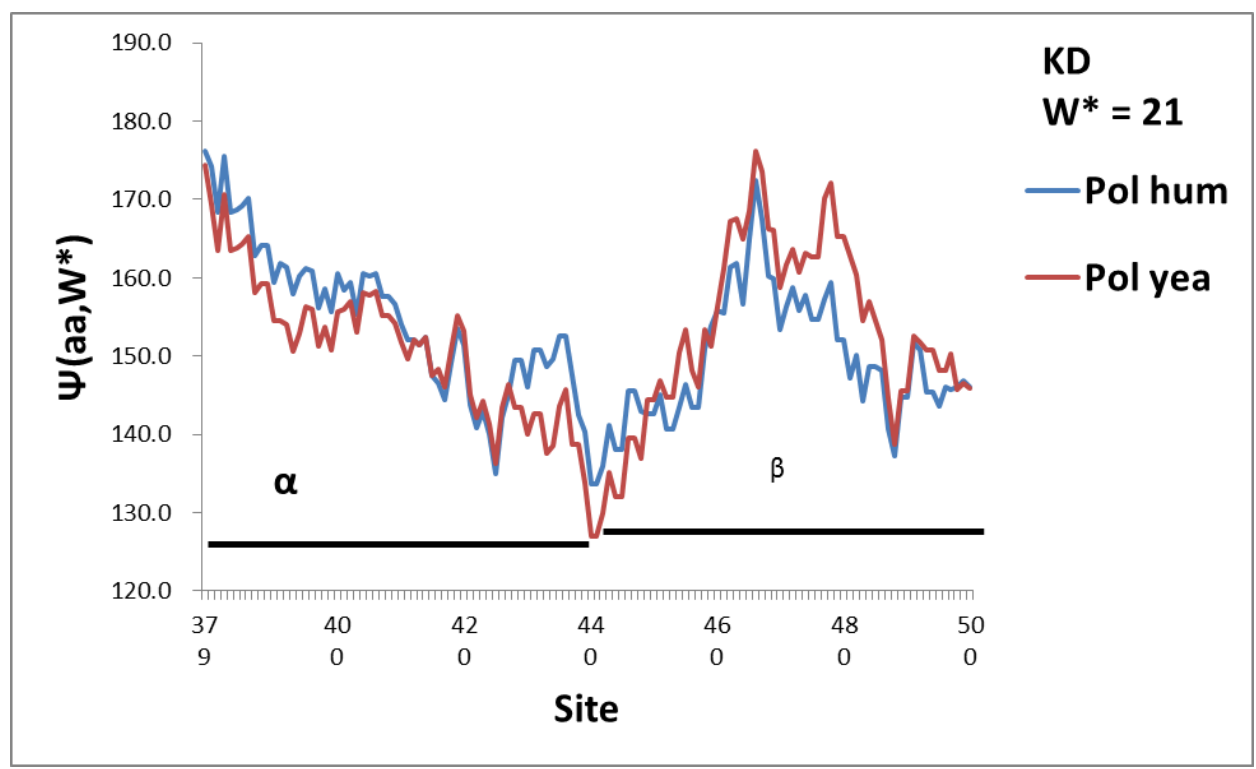

Fig. 5. A striking feature of the α + (reversed) β polymer profile is the deep hydrophilic minimum at the 440 connection. This deep minium is caused by acidic C terminal domain ends (see Figs. 1-4). Here we show an enlargement of the Cα - Cβ binding region profile. It is an asymmetric V, with broad and nearly linear amphiphilic sides. The Nα – Cβ binding region profile is similar but with a reversed V. Evolution has narrowed the human β profile, so that the ranges of the human binding regions' profiles are narrower, and their structures are more compact, than the yeast regions'. These differences are largest for the KD scale with W* = 21

.